

\overfullrule=0pt
\magnification=\magstep1
\baselineskip=3.3ex
\raggedbottom

\def\uprho{\raise1pt\hbox{$\rho$}}
\def\mfr#1/#2{\hbox{${{#1} \over {#2}}$}}
\def\1{{\bf 1}}
\def\Tr{{\rm Tr}}

\def\mod{{\rm mod}}
\def\rint{{\rm int}}
\def\R{{\cal R}}
\font\bcal=cmbsy10
\font\fivepoint=cmr5
\headline={\hfill{\fivepoint FLUX PHASE\ \ EHL14/Aug/94}}
\centerline{{\bcal THE \quad FLUX \quad PHASE \quad OF \quad THE \quad
HALF}{\bf -}{\bcal FILLED \quad BAND}}
\bigskip
\bigskip
\centerline{Elliott H. Lieb}
\centerline{Departments of Mathematics and Physics}
\centerline{Princeton University}
\centerline{P.O.B. 708, Princeton, NJ  08544}
\bigskip
\bigskip
{\narrower\smallskip\noindent
{\bf Abstract:}  The conjecture is verified that the optimum, energy
minimizing magnetic flux
for a half-filled band of electrons hopping on a planar, bipartite graph is
$\pi$ per square plaquette.  We require {\it only} that the graph has
periodicity in one direction and the result includes the hexagonal lattice
(with flux 0 per hexagon) as a special case.  The theorem goes beyond
previous conjectures in several ways:  (1) It does not assume, a-priori,
that all plaquettes have the same flux (as in Hofstadter's model); (2) A
Hubbard type on-site interaction of any sign, as well as certain longer
range interactions, can be included; (3) The
conclusion holds for positive temperature as well as the ground state; (4)
The results hold in $D \geq 2$ dimensions if there is periodicity in $D-1$
directions (e.g., the cubic lattice has the lowest energy if there is flux
$\pi$ in each square face). \smallskip}
\bigskip
\bigskip
\bigskip\noindent
PACS:  05.30.Fk, 75.10.Lp \hfill{For Physical Review Letters, {\bf 73},
2158 (1994)}
\bigskip
\bigskip
\bigskip

The flux phase conjecture states that the ground state (g.s.) energy minimizing
magnetic flux
through a square, planar lattice on which free electrons hop is $\pi$ per
plaquette when the electron filling factor is 1/2 [1-4].  (Zeeman terms are
excluded.)  This conjecture, along with
extensions to positive temperature, higher dimensional geometries and
allowance for some electron-electron interactions, will be proved here.

If the sites of the lattice are interpreted as atoms in a solid then
flux $\pi$ would correspond to magnetic fields available only on
neutron stars.  The significance of the flux phase is thus not
primarily as a literal interpretation in terms of physical magnetic
fields.  One interesting interpretation concerns mean field
calculations connected with superconductivity.  The main interest,
however, in the author's view, is that it shows that diamagnetism
(which states that the optimal flux is zero --- and which is correct
when the electron density is very small) can be reversed when the
density is high.  Indeed, it can be maximally reversed, as in this case
(since flux on a lattice is determined only modulo $2\pi$).  Thus,
there is a peculiar, non-intuitive and poorly understood effect of the
Pauli principle on the way in which orbital motion interacts with
magnetic fields.  It has been studied extensively [5-13]. See [10] for
some history.

To define things precisely, we start with a general {\it finite graph}
$\Lambda$, which is a collection of $\vert \Lambda \vert$ sites and
certain bonds denoted by $xy$  with $x$ and $y$ in $\Lambda$ and $x \neq y$.
A positive weight $\vert t_{xy} \vert = \vert t_{yx} \vert$ is
specified in advance for each bond.  By convention $t_{xx} = 0$.  The
hopping amplitude is then $t_{xy} = \vert t_{xy} \vert \exp [i \phi
(x,y)]$, with $\phi (x,y) = - \phi (y,x)$ for hermiticity, and the
problem is to find the numbers $\phi (x,y)$ that minimize the total
electronic ground state energy (when $\beta = 1/kT = \infty)$ or free
energy (when $\beta < \infty)$.

A {\it circuit} in $\Lambda$ is
a sequence of points $x_1, x_2, \dots , x_n, x_1$ with $t_{x_i
x_{i+1}} \not= 0$ for all $i$.  The flux through this circuit is $\sum
\nolimits^n_{i=1} \phi (x_i, x_{i+1}) \ (\mod 2 \pi)$.  It is a fact [10]
that the spectrum of the hermitian matrix $T = \{ t_{xy} \}_{x,y \in
\Lambda}$ depends on the $\phi$'s {\it only} through the fluxes.  This is
also true of the Hamiltonians below.
No a priori assumption is made that
the flux need be the same in all plaquettes; indeed, the flux is not
even assumed to be the same for up- and down-spin electrons.
We allow different $\vert t_{xy} \vert$'s and $\phi (x,y)$'s for the up- and
down-spin electrons.  We denote these by $T^\uparrow, T^\downarrow$.  Thus,
our results apply to the Falicov-Kimball model (where $T^\downarrow = 0$),
for example.

The {\it electronic kinetic energy operator}, in second-quantized notation, is
$$K = -\sum \limits_{x,y \in \Lambda} t^\uparrow_{xy} c^\dagger_{x \uparrow}
c^{\phantom{\dagger}}_{y \uparrow} + t^\downarrow_{xy} c^\dagger_{x\downarrow}
c^{\phantom{\dagger}}_{y \downarrow}. \eqno(1)$$
The $c$'s satisfy the fermion anticommutation relations
$\{ c^\dagger_{x \sigma}, c^{\phantom{\dagger}}_{y
\tau} \} = \delta_{xy} \delta_{\sigma \tau}, \ \{ c_{x \sigma}, c_{y \tau} \}
= 0.$ The
electron number is $N = N_\uparrow + N_\downarrow$ and $N_\sigma = \sum
\nolimits_{x \in \Lambda} n_{x \sigma}$ with $n_{x \sigma} = c^\dagger_{x
\sigma} c^{\phantom{\dagger}}_{x \sigma}$.  If the Hamiltonian $H$ equals
$K$, then the ground state energy $E_0$
would be $E_0 = \sum \nolimits_{\lambda < 0} \lambda (T^\uparrow) +
\sum \nolimits_{\lambda < 0} \lambda (T^\downarrow)$, i.e., the sum of
the negative eigenvalues, $\lambda (T)$, of the matrices $T^\uparrow$
and $T^\downarrow$.

For {\it bipartite graphs} (i.e., $\Lambda = A \cup B, \ A \cap B =
\emptyset$ and $t_{xy} = 0$ unless $x \in A, \ y \in B$ or $x \in B, \ y
\in A$) $E_0$ is achieved when $N = \vert \Lambda \vert$, and hence the
appelation half filled band, since $0 \leq N \leq 2 \vert \Lambda \vert$ in
general.
In the Hubbard model $H = K + W^0$,
$$W^0 = \sum \limits_{x \in \Lambda} U_x (n_{x \uparrow} - \mfr1/2) (n_{x
\downarrow} - \mfr1/2), \eqno(2)$$
but we can also add
certain longer range density-density interactions, $W^d$ and spin-spin
interactions $W^s$ to be specified
later, of the form (with $w^k_{xy} = w^{k*}_{xy} =
w^k_{yx}$, $k = d, 1, 2, 3$ and $S^j$ being Pauli matrices)
$$\eqalignno{W^d &= \sum \limits_{x,y \in \Lambda} w^d_{xy} (n_{x \uparrow}
+ n_{x \downarrow} - 1) (n_{y \uparrow} + n_{y \downarrow} - 1) \cr
W^s &= \sum \limits^3_{j=1} \sum \limits_{x,y \in \Lambda} \sum
\limits_{\sigma, \tau, \mu, \lambda} w^j_{xy}
(c^\dagger_{x \sigma}  S^j_{\sigma \tau} c^{{\phantom{\dagger}}}_{x \tau})
(c^\dagger_{y \mu}  S^j_{\mu \lambda} c^{{\phantom{\dagger}}}_{y \lambda}).
\qquad&(3)\cr}$$

$W^0, W^d$ and $W^s$ are invariant under the unitary hole-particle
(h-p) transformation $\tau$, with $\tau
c^{\phantom{\dagger}}_{x \sigma} \tau^{-1} = c^\dagger_{x \sigma}$.
The kinetic energy operator with complex $T$'s satisfies
$$\tau K (T^\uparrow, T^\downarrow) \tau^{-1} = K (-T^{\uparrow *},
-T^{\downarrow *} )\eqno(4)$$
where $T^*$ denotes the complex conjugate matrix $t^{\phantom{*}}_{xy}
\rightarrow t^*_{xy} = t^{\phantom{*}}_{yx}$.

The grand-canonical partition function of our system, which we want to
maximize, is $Z = \Tr \exp [-\beta H]$ at inverse temperature $\beta$.  By
h-p symmetry, $\langle N \rangle \equiv \Tr Ne^{-\beta H}/Z$ is $\vert \Lambda
\vert$, which is the half-filled band.  The $\beta \rightarrow \infty$
limit is discussed at the end.

Henceforth, all graphs will be bipartite, in which case all elementary
circuits contain an even number of sites and bonds.  The original flux
phase conjecture is that when $H = K, \ N = \vert \Lambda \vert$ and
$\Lambda$ is planar the optimum choice of fluxes is $\pi$ in every
circuit containing $0(\mod 4)$ sites and 0 in circuits with $2 (\mod
4)$ sites.

Several cases of this were proved in [9, 10] and it was pointed out in [10]
that the conjecture can not always hold for {\it arbitrary} values of
$\vert t_{xy} \vert$.  It depends on $\Lambda$.
Despite this caveat, however, it was proved in [10] that log [det$(T^2)]
= \sum \nolimits_j \log [\lambda_j (T)^2]$ is maximized when the fluxes
accord with the conjecture.  This is true for an {\it arbitrary} bipartite,
planar graph with {\it arbitrary} $\vert t_{xy} \vert$'s.  (On any bipartite
graph, the nonzero eigenvalues of $T$ come in opposite pairs $\lambda,
-\lambda$.)

The {\it generalized flux phase conjecture} is that the above choice is also
optimal for $H = K + W^0 + W^d + W^s$ and for all $\beta \leq \infty$.
We shall
prove this here for graphs that have a certain periodicity. The usual
square lattice with periodic boundary conditions is included. The result
holds also for higher dimensional, non-planar graphs.
The type of graphs $\Lambda$ considered here is illustrated in
Fig. 1 for the planar case.  $\Lambda$ is wrapped on a cylinder,
i.e., the sites at the right end are identified with the sites on the
left.  The $\vert t_{xy} \vert$'s on the vertical edges must be periodic,
i.e., they are allowed to vary in an arbitrary way along each column, but
all the columns must be identical.  The horizontal $\vert t_{xy} \vert$'s
can also vary as we move vertically but they are only required to have
period 2 in the horizontal direction, i.e., every {\it second} column of
horizontal edges must be the same.  Thus, if we erase the horizontal edges
1, 3, 5, etc. from every second column and if we erase edges 2, 4, 6, etc.
from the remaining columns in between, the hexagonal lattice is obtained.
Our result includes this case, and flux $\pi$ in each (imaginary) square then
implies flux 0 in each hexagon.  Octagons, decagons, etc.  can also
be included provided periodicity 2 is
maintained.  We can, if we wish, insert vertical edges connecting the
bottom row to the top row, as indicated by the vertical arrows in Fig. 1.
$\Lambda$ will no longer be planar, but that
does not matter.

An easier way to state the periodicity requirements is to cut $\Lambda$
at the two dashed lines, called P in Fig. 1.  The two half-cylinders
(as well as the $\vert t_{xy} \vert$'s on the edges) are required to be
mirror images of each other.  The $\vert t_{xy} \vert$'s on the edges
that intersect the dashed lines are arbitrary because each is its own
mirror image.  We then say that $\vert T \vert$ is {\it reflection
symmetric} with respect to the cutting lines, P, through the bonds.
Our theorem says that when $H=K$ the optimum flux is then $\pi$ for
those squares containing the cutting lines.  If the $\vert t_{xy}\vert$'s
are reflection symmetric w.r.t. {\it every} choice of cutting lines --
which is equivalent to the above periodicity requirement -- then flux
$\pi$ will be optimal in {\it every} square of $\Lambda$.

The periodicity of the $\vert t_{xy} \vert$'s mentioned above is {\it not}
needed for the theorem below.  Only reflection symmetry is needed.  The
periodicity comes in when we wish to insure flux $\pi$ in {\it every}
plaquette of $\Lambda$.  This is achieved by repeated reflection in
hyperplanes in the standard way [14-15]; indeed, one can easily derive the
usual chessboard estimates [14].

In fact $\Lambda$ could be built in a similar way out of $D$-dimensional
(hyper)cubes instead of
squares.  Cutting lines become cutting ($D-1$)-dimensional hyperplanes, and
reflection symmetry is generalized in an obvious way.  Our theorem will then
state that the optimal flux in each two-dimensional square
plaquette of the (hyper)cubic lattice is $\pi$ in every plaquette cut by
the hyperplanes.  As in the $D=2$ case, flux $\pi$ will be optimal in {\it
every} plaquette if we have periodicity in $D-1$ directions.

Returning to the two-dimensional situation (with obvious generalization
to $D > 2$) we require only that the $U_x$'s in $W^0$ be constant in
the horizontal direction (as are the $\vert t_{xy} \vert$'s on vertical
edges).  As for $w^k_{xy}$ in $W^d$ and $W^s$, they are required to be
reflection positive for reflection in vertical planes (the dashed
lines, $P$, in Fig. 1), as explained below.  The inclusion of $W^d$ and
$W^s$ is mainly for completeness and nothing essential will be lost by
setting $W^d = W^s = 0$.

In general, a Hamiltonian, $H$, can be written (with respect to a cutting
hyperplane, P) as
$$H = H_L + H_R + H_{\rint} \eqno(5)$$
where $H_L$ is all the terms involving sites in the left half-cylinder,
$H_R$ involves right half-cylinder sites and $H_{\rint}$ involves both.
The subscripts $L, R$, and $int$ will also be used for the separate pieces,
e.g.
$K_L, K_R, K_{\rint}$, etc.
Associated with the left Hamiltonian $H_L$ is a right Hamiltonian, $\Theta
(H_L)$, which is obtained from $H_L$ by
three steps:  {\it the unitary transformation $\R$ generated by geometric
reflection through the plane $P$; hole-particle transformation
$\tau$; and complex conjugation $*$}.  $\Theta (H_L) = (\tau \R (H_L)
\tau^{-1})^*$.  This notion of {\it
operator reflection} can be applied to
any operator, $A_L$ in the left algebra (i.e., $A_L$ is a polynomial in the
operators $c^{{\phantom{\dagger}}}_{x \sigma}, c^\dagger_{x \sigma}$ with
$x$ in the left half-cylinder).  In particular $\theta
(c^{{\phantom{\dagger}}}_{l\sigma}) = c^\dagger_{r \sigma}$.
A similar definition holds for the left
Hamiltonian $\Theta (H_R)$ and, clearly, $\Theta (\Theta (H_L)) = H_L$.
Note, from (4), that $\Theta (K_L) = - \R (K_L), \ \Theta (W^\alpha) = \R
(W^\alpha), \ \alpha = 0, d, s$.

{\it Reflection positivity} of $w_{xy}$ means that it is
symmetric under reflections and, for $x$ in
the left and $y$ in the right half-cylinder,
$w_{xy}$ can be written as a sum (or integral) of functions of the form
$a(x) a (\R y)^*$.  In other words, $W^d_{\rint}, W^s_{\rint}$ are sums (or
integral) of operators of the form $-A_L \Theta (A_L)$, where the $(-)$ in
$W^d_{\rint}$
comes from $\tau (n_{x\uparrow} + n_{x \downarrow} - 1) \tau^{-1} = 1 -
n_{x \uparrow} - n_{x \downarrow}$, and similarly for $W^s_{\rint}$.
An important example is
$w_{xy} = w > 0$ if $x,y$ are nearest neighbors, $w_{xy} =0$ otherwise.
Such a $W^s$ is antiferromagnetic.  See [14].

Concerning $K_{\rint}$, we note that it is generally not invariant
under the three operations.  However, with $l$ and $r$ denoting a
generic left, right image pair cut by $P$, we are at liberty to choose
$\Theta (l,r) = 0$, i.e., $t_{lr} = \vert t_{lr} \vert \geq 0$, and we
do so.  (Note:  to simplify the notation the symbols, $\uparrow$ and
$\sigma$, will not be indicated.) This is so because a simple gauge
transformation $c_r \rightarrow \exp [-i \Theta (l,r)] c_r$ makes
$t_{lr} > 0$ without changing any fluxes.  No circuits are involved.
This choice of phase for $t_{lr}$ is only a convention, for it does not
change any physics, but it is important for (6) and (7) below.

With the foregoing convention for $H_{\rint}$, the Hamiltonian is said to
be {\it reflection symmetric} if $\Theta (H_L) = H_R$.
The flux $\pi$ theorem will be a corollary of the following lemma.

{\bf LEMMA (Reflection positivity).}  {\it With $H$ as given in (5) with
respect to some hyperplane, $P$, assume that $K_{\rint}$ satisfies the
above positivity convention.  Assume also that $W^d_{\rint}$ and
$W^s_{\rint}$ are reflection
positive.  Then, for each $\beta \geq 0$ and with $H_{\rint}$ fixed,
$$Z (H_L, H_R)^2 \leq Z(H_L, \Theta (H_L)) Z(\Theta (H_R), H_R), \eqno(6)$$
where $Z(H_L, H_R) \equiv \Tr \exp [-\beta H]$.  Moreover, if $H_R = \Theta
(H_L)$ and if $A_L$ (resp. $A_R$) is any even operator in the left (resp.
right) algebra (e.g., $A_L$
is a sum of monomials in $c^\#_{x \sigma}$ of even degree) then
$$\vert \Tr A_L A_R e^{-\beta H} \vert^2 \leq \Tr A_L \Theta (A_L)
e^{-\beta H} \,\Tr \Theta (A_R) A_R e^{-\beta H}\quad. \eqno(7)$$}

{\it Proof:}
Use the Lie-Trotter formula to approximate $e^{-\beta H}$ as a
product of $M \gg 1$ factors $V = V_{\rint} V_L V_R$, i.e., $e^{-\beta H} =
\lim \nolimits_{M \rightarrow \infty} V^M$,
where $V_{\rint} = (1 - \beta H_{\rint}/M)$, $V_L = \exp [-\beta H_L/M], V_R =
\exp [-\beta H_R/M]$.  Notice that $V_L$ contains only even polynomials in
the $c^\#$'s, and so $V_L$ commutes with every right operator
(including odd operators).  Likewise, $V_R$ commutes with all left operators.

If the $M$ factors of $V_{\rint}$ are multiplied out we obtain for $V^M$
a sum of terms, each having the form
$X = a_1 V_L V_R a_2 V_L V_R a_3 V_L V_R \cdots a_M V_L V_R$
and each $a_i$ has one of three forms:  (i) $A_L \Theta (A_L)$, with $A_L$
an even operator or (ii) $c^\dagger_l c^{{\phantom{\dagger}}}_r$ or (iii)
$-c^{{\phantom{\dagger}}}_l c^\dagger_r$.  Our
strategy is to move all the left operators to the left {\it without}
changing the order either of the left operators among themselves or the
right operators.  The operators $A_L$ commute with all the right operators
and cause no difficulty.  The difficult point is that the $c^\#_l$ operators
have to move through the $c^\#_r$ operators to their left, and each such
move gives rise to a $-1$ factor.

I claim that either $\Tr X = 0$ or else the number of $-1$ factors is
even.  To see this note that by particle conservation (and the particle
conserving nature of $V_L$ and $V_R$) the number of $c^\dagger_l
c^{{\phantom{\dagger}}}_r$ factors must equal the number of
$c^{{\phantom{\dagger}}}_l c^\dagger_r$ factors if $\Tr X
\not= 0$.  Call this common number $J$.  The number of $-1$ factors is
independent of the order of these $2J$ factors and their
order relative to the $A_L \Theta (A_L)$
factors.  The first $c^\#_l$ must move through zero $c^\#_r$'s.  The
second $c^\#_l$ moves through one $c^\#_r$, etc..  Thus, the number of
$-1$ factors is $0 + 1 + 2 + \cdots + (2J-1) = J(2J-1)$.  On the other
hand, each $c^{{\phantom{\dagger}}}_l c^\dagger_r$ term carries a $-1$
factor and there
are $J$ of these.  Altogether there are $J + J (2J-1) = 2J^2 = 0(\mod 2)$
factors of $-1$, as claimed.

In brief, $X$ can be brought into the form $X = X_L X_R$ with $X_L$ and
$X_R$ even operators.  Since $\Tr \1 = 4^{\vert \Lambda \vert}$, we have
$4^{\vert \Lambda \vert} \Tr X = \Tr X_L \Tr
X_R$.  Moreover, $(\Tr X_L)^* = \Tr \Theta (X_L)$ and
thus $\vert \Tr X_L \vert^2 = \Tr X_L \Theta (X_L)$.  Now, denoting the
various $X$'s by $X^\alpha$, we have $\vert \Tr V^M \vert^2 = \vert \sum
\nolimits_\alpha \Tr X^\alpha \vert^2 = 4^{-2 \vert \Lambda \vert} \vert \sum
\nolimits_\alpha \Tr X^\alpha_L \Tr X^\alpha_R \vert^2 \leq 4^{-2 \vert \Lambda
\vert} \sum \nolimits_\alpha \vert \Tr X^\alpha_L \vert^2 \sum
\nolimits_\alpha \vert \Tr X^\alpha_R \vert^2 = \sum \limits_\alpha \Tr
X^\alpha_L \Theta (X^\alpha_L) \sum \nolimits_\alpha \Tr X^\alpha_R \Theta
(X^\alpha_R) \rightarrow Z (H_L, \Theta (H_L)) Z (H_R, \Theta (H_R))$.
(7) is obtained in the same way.  QED

{\bf THEOREM (Flux $\pi$ is optimal).}  {\it Assume the $\vert t^{\uparrow,
\downarrow}_{xy} \vert$ are reflection invariant w.r.t. $P$.  Assume also
that $\Theta (W^\alpha_L) = W^\alpha_R$ and $W^\alpha_{\rint}$ is
reflection positive, $\alpha = 0, d, s$.  Then $Z$ is maximized by putting
flux $\pi$ in each square face of $\Lambda$ that intersects $P$.}

{\it Proof:}  We make the gauge transformation above so that $K_{\rint}$
has $t_{lr} = \vert t_{lr} \vert$.  From (6), we have that when $H_L, H_R$
is optimal, so is $H_L, \Theta (H_L)$ and $\Theta (H_R), H_R$.  But the
statement $K_R = \Theta (K_L)$ implies the flux $\pi$ condition by (4).  QED

{\it Remarks:}  (i).  It is interesting to note that (6) can also be used
to show that when the fluxes are fixed at $\pi$ and one varies over the
$\vert t_{xy} \vert$, the lowest energy is attained in a reflection
symmetric configuration of $\vert t_{xy} \vert$.  A one-dimensional version
of the lemma was proved and used in [16] to study the Peierls instability
for the Hubbard model on a ring.

(ii).  The lemma and theorem say that flux
$\pi$ for $T^\uparrow$ and $T^\downarrow$ is optimal.  If we fix
$\phi^\uparrow (x,y)$, we are then free to choose $\phi^\downarrow (x,y) =
\phi^\uparrow (x,y)$, since any other choice with flux $\pi$ differs from
$\phi^\uparrow (x,y)$ by a trivial gauge transformation, $c_{x \downarrow}
\rightarrow e^{i \mu (x)} c_{x \downarrow}$.  Thus, when $\vert
t^\uparrow_{xy} \vert = \vert t^\downarrow_{xy} \vert$, the minimizer can
have $T^\downarrow = T^\uparrow$, thereby preserving $SU(2)$ invariance.

(iii).  To discuss the ground state we let $\beta \rightarrow \infty$.
Do we get $N = \vert \Lambda \vert$ or does the g.s.  belong to $N =
\vert \Lambda \vert + m$ and $N = \vert \Lambda \vert - m$ with $m >
0$?  In the Falicov-Kimball model, generally, the g.s. has $N = 2 \vert
A \vert$ and $N = 2 \vert B \vert$, as the only choices [17] (but note
that $2 \vert A \vert = 2 \vert B \vert = \vert \Lambda \vert$ in our
case).  In the following cases I can also prove that at least one g.s.
has $N = \vert \Lambda \vert$.

First, assume reflection symmetry and positivity w.r.t. {\it all}
hyperplanes parallel to $P$, so that we are now looking at a $K$ with flux
$\pi$ in {\it every} plaquette of $\Lambda$.  After a trivial gauge
transformation, this condition can be realized with real $T$, which we
assume henceforth.  Next, assume $T^\uparrow = T^\downarrow$, so that $SU(2)$
invariance holds. Third, assume $W^d = W^s = 0$, i.e., the Hubbard model.
As is well known, we can then
construct another set of $SU(2)$ generators --- the {\it pseudospin}.  (See
[18] and, for more details, [19].)  By using the spin and pseudospin
raising operators $\sum \nolimits_x c^\dagger_{x \uparrow}
c^{{\phantom{\dagger}}}_{x \downarrow}, \  \sum \nolimits_x (-1)^x
c^\dagger_{x \uparrow} c^\dagger_{x \downarrow}$, (with $(-1)^x = + 1$ for
$x \in A$, $-1$ for $x \in B$) and their adjoints, one can conclude that
the absolute ground state belongs to $N = \vert \Lambda \vert$ or $N =
\vert \Lambda \vert \pm 1$.

Finally, to show that the g.s. has $N = \vert \Lambda \vert$, assume either
that $U_x \leq 0$ for all $x$ or $U_x \geq 0$ for all $x$.  We can then use
spin-space reflection positivity [17] in {\it Fock space}, together with
the evenness of $\vert \Lambda \vert$ in our case, to infer $N = \vert
\Lambda \vert$.  (This reflection positivity tells us that if a g.s. has
numbers $N_\uparrow = \lambda, N_\downarrow = \mu$ then there are ground
states with $(\lambda, \lambda)$ and $(\mu, \mu)$.  Thus, if $\vert \Lambda
\vert = 2m$ and $N = \vert \Lambda \vert - 1$, so that $N_\uparrow = m, \
N_\downarrow = m - 1$, then there is also an $(m,m)$ g.s..)  If all $U_x
\not= 0$, the g.s. is unique [18].

{\it Extensions:}  The flux phase for the half-filled band has been proved
here for a large class of Hamiltonians, including the ones common in the
physics literature.  The proof is sufficiently simple that
it obviously applies to many other models.

One generalization is to fermions with $n \neq 2$ colors, i.e. from $SU(2)$ to
$SU(n)$.

Certain specialized forms of electron-phonon interactions can be included.

Another generalization is to $SU(2)$ instead of $U(1)$ gauge fields [15,
20].  Thus, $t_{xy} \sum \nolimits_\sigma c^\dagger_{x \sigma}
c^{{\phantom{\dagger}}}_{y \sigma}$ is replaced by $\vert t_{xy} \vert
\sum \nolimits_{\sigma \tau} c^\dagger_{x \sigma} U^{\sigma \tau}_{xy}
c^{{\phantom{\dagger}}}_{y \tau}$ with $\vert t_{xy} \vert$ given, as
before, and with $U_{xy} \in SU(2)$ to be determined.  (Even more
generally, we can replace $t^\uparrow_{xy} c^\dagger_{x \uparrow}
c^{{\phantom{\dagger}}}_{y \uparrow} + t^\downarrow_{xy} c^\dagger_{x
\downarrow} c^{{\phantom{\dagger}}}_{y \downarrow}$ by $\sum
\nolimits_{\sigma \lambda \tau} c^\dagger_{x \sigma} M^{\sigma
\lambda}_{x; xy} \vert t^\lambda_{xy} \vert M^{\lambda \tau}_{y;xy}
c^{{\phantom{\dagger}}}_{y \tau}$ where $M_{x;xy} \in SU(2)$ and $\vert
t^\lambda_{xy} \vert$ is given; in this case the $SU(2)$ matrix
associated with $xy$ is $U_{xy} = M_{x;xy} M_{y;xy}$.)  Again, we will
find that the energy is minimized by flux $\pi$ in each plaquette,
i.e., the product of the four matrices around a plaquette satisfies
$U_{xy} U_{yz} U_{zw} U_{wx} = - 1$.

Thanks are due to I.~Affleck, V.~Bach, J.~Bellissard, E.~Carlen, J.~Fr\"ohlich,
M.~Loss, J.B.~Marston, B.~Nachtergaele, J.P.~Solovej and P.~Wiegmann for
helpful
discussions and to the U.S. National Science Foundation, grant
PHY90-19433-A03, for partial support.

\noindent
{\bf REFERENCES}
\item{[1]}  I. Affleck and J.B. Marston,
Phys. Rev. {\bf B37}, 3774 (1988).
\item{[2]}  Y. Hasegawa, P. Lederer, T.M. Rice and P.B. Wiegmann,
Phys. Rev. Lett. {\bf 63}, 907 (1989).
\item{[3]}  G. Kotliar, Phys. Rev. {\bf B37}, 3664 (1988).
\item{[4]}  D.S. Rokhsar, Phys. Rev. {\bf B42}, 2526 (1990);
Phys. Rev. Lett. {\bf 65}, 1506 (1990).
\item{[5]}  A. Barelli, J. Bellissard and R. Rammal,
J. Phys. (France) {\bf 51}, 2167 (1990).
\item{[6]}  J. Bellissard and R. Rammal, J. Phys. (France) {\bf
51}, 1803 (1990); J. Phys. (France) {\bf
51}, 2153 (1990); Europhys. Lett. (Switzerland) {\bf 13}, 205 (1990).
\item{[7]}  D.R. Hofstadter, Phys. Rev. {\bf B14}, 2239 (1976).
\item{[8]}  P.G. Harper, Proc. Phys. Soc. Lond. {\bf A68}, 874
(1955); Proc. Phys. Soc. Lond. {\bf 68A}, 879 (1955).
\item{[9]}  E.H. Lieb, Helv. Phys. Acta {\bf 65}, 247 (1992).
\item{[10]}  E.H. Lieb and M. Loss, Duke Math. Jour. {\bf 71}, 337 (1993).
\item{[11]}  P.B. Wiegmann, Physica {\bf 153C}, 103 (1988).
\item{[12]}  X.G. Wen, F. Wilczek and A. Zee, Phys. Rev. {\bf B39}, 11413
(1989).
\item{[13]} F. Nori and Y-L. Lin, Phys. Rev. {\bf B49}, 4131 (1994). See
also F. Nori, B. Dou\c cot and R. Rammal, Phys. Rev. {\bf B44}, 7637; F.
Nori, E. Abrahams and G.T. Zimanyi, Phys. Rev. {\bf B41}, 7277 (1990).
\item{[14]}  J. Fr\"ohlich, R. Israel, B. Simon and E. Lieb,
Commun. Math. Phys. {\bf 62}, 1 (1978).
\item{[15]}  D. Brydges, J. Fr\"ohlich and E. Seiler,
Ann. Phys. (NY) {\bf 121}, 227 (1979).
\item{[16]}  E.H. Lieb and B. Nachtergaele, {\it The stability of the
Peierls instability for ring-shaped molecules}, in preparation.
\item{[17]}  T. Kennedy and E.H. Lieb, Physica {\bf 138A}, 320 (1986).
\item{[18]}  E.H. Lieb, Phys. Rev. Lett. {\bf 62}, 1201 (1989); Errata 1927
(1989).
\item{[19]}  E.H. Lieb, {\it The Hubbard model: Some rigorous results and
open problems}, in Proceedings of 1993 conference in honor of G.F.
Dell'Antonio, ``Advances in dynamical systems and quantum physics'', World
Scientific (in press) and in Proceedings of 1993 NATO ASW ``The physics and
mathematical physics of the Hubbard model'', Plenum (in press).
\item{[20]}  Y. Meir, Y. Gefen and O. Entin-Wohlman, Phys. Rev. Lett. {\bf
63}, 798 (1989).
\bigskip
\bigskip
\bigskip

\centerline{{\bf FIGURE CAPTION}}

{\bf FIG. 1.}  Typical 2D lattice with horizontal periodic boundary
conditions (left boxes $=$ right boxes).  Different bond weights
illustrate
the requirement of horizontal periodicity 2.  Dashed lines (P) are a
reflection plane.  A generic left-right pair of sites is indicated by
$l,r$.

\bye